\documentstyle[aps,preprint,tighten,epsfig]{revtex}
\begin{document}
\title{ Extracting the Pomeron-Pomeron Cross Section from Diffractive Mass Spectra} 
\author{F.O. Dur\~aes$^1$\thanks{e-mail: fduraes@if.usp.br}, \ F.S.
Navarra$^{1,2}$\thanks{e-mail: navarra@if.usp.br} \ and \ G.
Wilk$^{2}$\thanks{e-mail: wilk@fuw.edu.pl} \\[0.1cm]
{\it $^1$Instituto de F\'{\i}sica, Universidade de S\~{a}o Paulo}\\
{\it C.P. 66318,  05389-970 S\~{a}o Paulo, SP, Brazil} \\[0.1cm]
{\it$^2$Soltan Institute for Nuclear Studies, 
Nuclear Theory Department (Zd-PVIII)}\\
{\it ul. Ho\.za 69, \ 00-681 Warsaw, Poland}}
\maketitle
\begin{abstract}
We have calculated the mass distribution ($d\sigma/dM_X$) as observed by UA8 Collaboration 
in the inclusive reaction  $p \bar p \rightarrow p X \bar p$ at $\sqrt s = 630  \,GeV$, using the Interacting 
Gluon Model (IGM) with Double Pomeron Exchange (DPE) included. The only new parameter is the $I\!\!P$-$I\!\!P$ 
cross section, which we can extract from fitting experimental data. We compare our results with the values obtained  
in the UA8 study.  Assuming a constant  Pomeron-Pomeron total cross section ($\sigma_{I\!\!P  I\!\!P} = 1 \, mb$), 
we make predictions for $d\sigma/dM_X$ at Tevatron and LHC energies.\\


\end{abstract}

\section{Introduction}

After ten years of work at HERA, an impressive amount of knowledge about the Pomeron has been accumulated, 
especially about its partonic composition and parton distribution functions. Less known are its interaction 
properties. Whereas the Pomeron-nucleon cross section has been often discussed in the literature, the recently 
published data by the  UA8 Collaboration \cite{UA8} have shed some light on the  Pomeron-Pomeron  interaction.  
In  \cite{UA8} the Double Pomeron Exchange (DPE) cross section was written  as the product of two 
flux factors with the   $I\!\!P$-$I\!\!P$ cross section, $\sigma_{I\!\!P I\!\!P}$, being thus directly 
proportional to this quantity. This simple formula relies on the validity of the  Triple-Regge 
model, on the universality of the Pomeron flux factor and on the existence of a factorization formula for
DPE processes. However, for these processes the factorization hypothesis has not been 
proven and is still matter of debate \cite{collins,bersop,bercol,terron}. In \cite{berera} it was shown that 
factorizing and  non-factorizing DPE models may be experimentally distinguished in the case of dijet production.

Fitting the measured  mass spectra allowed for the determination of $\sigma_{I\!\!P I\!\!P}$ 
and its dependence on $M_X$, the mass of the diffractive system. The first observation of the UA8 analysis was 
that the measured diffractive mass ($M_X$) spectra show an excess at low values  that can hardly be explained 
with a constant (i.e., independent of $M_X$)  $\sigma_{I\!\!P I\!\!P}$. Even after introducing some mass 
dependence in  $\sigma_{I\!\!P I\!\!P}$ they were not able to  fit the spectra in a satisfactory way.  Their 
conclusion was that the low $M_X$ excess may have some physical origin like, for example glueball formation.

Although the  analysis performed in \cite{UA8}  is standard, it is nevertheless useful to confront it with 
other, also successful, descriptions of the diffractive interaction. One of them is the one provided by the 
Interacting Gluon Model (IGM)\cite{IGM}. This model describes only certain aspects of hadronic collisions, 
related to energy flow and energy deposition in the central rapidity region. It should not be regarded as an 
alternative to a field-theoretical approach to diffractive amplitudes, but rather as an extension of the naive 
parton model.  The reason for using it here is that it may be good enough to account for energy flow in an 
economic way. The deeper or more subtle aspects of the underlying field theory probably (this is our belief) 
do not manifest themselves in energy flow, but rather in other quantities like the total cross section. 
Inspite of its simplicity, this model can teach us a few things and predict another few. 
This is encouraging 
because in the near future new data about DPE from CDF will be available \cite{gou}. 
In this work we would like to address the UA8 data  with the IGM.  
As it will be seen, according to our analysis the low mass behavior of the $M_X$ spectra  can be explained 
with a constant $I\!\!P$-$I\!\!P$ cross section.

\section{The IGM}

The IGM has been described at length, especially in \cite{IGM} and more recently in \cite{SYST}. 
In the past
we have successfully modified the IGM in such way as to include in it hadronic single diffractive dissociation 
processes \cite{IGM97,hera,SYST,LEADING} and applied it to hadronic collisions and HERA-photoproduction data 
($M_X^2$ distributions and leading particles spectra of $p$, $\pi$, $K$ and $J/\psi$).

The main idea of the model is that nucleon-nucleon collisions at high 
energies can be treated as an incoherent sum of multiple gluon-gluon collisions, the valence quarks playing 
a secondary role in particle production. While this idea is well  accepted for large momentum transfer between 
the colliding partons, being on the basis of some models of minijet and jet production (for example
HIJING \cite{wang}), in the IGM 
its validity is extended down to low momentum transfers, only slightly larger than $\Lambda_{QCD}$. At first 
sight this is not  justified because at lower scales there are no independent gluons, but rather a highly 
correlated configuration of color fields. There are, however, some indications coming from lattice QCD calculations, 
that these soft gluon modes are not so strongly correlated. One of them is the result obtained in \cite{pana}, namely 
that the typical correlation length of the soft gluon fields is close to $0.3\,fm$. Since this length is still much 
smaller than the typical hadron size, the gluon fields can, in a first approximation, be treated as uncorrelated.
Another independent result
concerns the determination of the typical instanton size in the QCD vaccum, which turns out to be of the order 
of $0.3\,fm$ \cite{sha}. As it is well known (and has been recently applied to high energy nucleon-nucleon and 
nucleus-nucleus
collisions) instantons are very important as mediators of  soft gluon interactions \cite{shu}. The 
small size of the active instantons lead to short distance  interactions between soft gluons, which can be treated 
as independent. 

These two results 
taken together suggest that a collision between the two gluon clouds (surrounding the valence quarks) may be viewed
as a sum of independent binary gluon-gluon collisions, which is the basic idea of our model.   
Developing the  picture above with standard techniques and enforcing  energy-momentum conservation, the IGM becomes 
the ideal tool to study energy flow in high energy hadronic collisions, in particular leading particle production and 
energy deposition. Confronting this simple model with  several and different data sets we obtained surprisingly 
good agreement with  experiment.

As indicated in the recent literature \cite{collins,bersop,bercol,terron,berera}, one of the crucial issues in 
diffractive physics is the possible breakdown of factorization. As stated in \cite{bersop} one may have Regge 
and hard factorization. Our model does not rely on any of them.  In the language used in \cite{bersop}, we 
need and use a ``diffractive parton distribution'' and we do not really need to talk about
``flux factor'' or ``distribution of partons  in the Pomeron''. Therefore there is no Regge 
factorization implied. However, we will do this connection in Eq. (\ref{conv}), in order to make contact   
with the Pomeron pdf's parametrized by the H1 and ZEUS collaborations.
As for hard factorization, it is valid as long as the 
scale $\mu$ is large. In the IGM, as it will be seen,  the scale is given by  $\mu^2 = x y s$, a number which 
sometimes is larger than $3 - 4 \, GeV^2$ but sometimes is smaller, going down to values only slightly
above $\Lambda_{QCD}^2$. When the scale is large ($ \mu^2 > p_{T_{mim}}^2$)  we employ Eq. (\ref{eq:OMEGAH}) 
and  when it is smaller ($m^2_0 < \mu^2 < p_{T_{mim}}^2$) we use  Eq. (\ref{eq:OMEGAS}).
Therefore, in part of the phase space we are inside the validity domain of hard  
factorization, but very often we are outside this domain. From the practical point of view, 
Eq. (\ref{eq:OMEGAH}), being defined at a semihard scale, relies on hard factorization for 
the elementary $gg \rightarrow gg$ interaction,  uses parton distribution function extracted from DIS 
and an elementary cross section $\hat{\sigma}_{gg}$ taken from standard pQCD calculations. 
The validity of the factorizing-like formula Eq. (\ref{eq:OMEGAS}) of our paper is {\it an assumption of the model}. 
In fact, the relevant scale there is $m_0^2 \simeq \Lambda_{QCD}^2$ and, strictly speaking,  there are no
rigorously defined parton distributions, neither elementary cross sections. However, using  Eq. (\ref{eq:OMEGAS})
has non-trivial consequences which were in the past years supported by an extensive comparison with 
experimental data.

\section{Double Pomeron Exchange}

Double Pomeron exchange  processes, inspite of their small cross sections,  appear to be
an excellent testing ground for the IGM because they are inclusive measurements and do not involve particle 
identification, dealing only with energy flow. In what follows, we briefly mention our main formulas. For further 
discussion we refer to the works \cite{IGM97,hera,SYST}. 

\subsection{Kinematics}

In Fig. 1 we show schematically the IGM picture of a double Pomeron exchange event in a proton-antiproton 
collision. The interaction follows then the usual IGM  \cite{IGM} picture, namely: the valence quarks fly through 
essentially undisturbed whereas the gluonic clouds of both projectiles interact strongly with each other (by gluonic 
clouds we understand a sort of ``effective gluons" which include also their fluctuations seen as 
$\bar{q}q$ sea  pairs). 
The proton (antiproton) looses fraction $x$ ($y$) of its original momentum and gets excited 
forming what we call a leading jet carrying $x_p= 1 -x$ ($x_{\bar p}= 1 -y$) fraction of the initial 
momentum.

According to the IGM \cite{SYST},  the probability to form a fireball carrying momentum fractions $x$ and $y$ of two 
colliding hadrons (see Fig. 1) is given by: 
\begin{eqnarray}
\chi (x,y) &=&\frac{\chi _0}{2\pi \sqrt{D_{xy}}} \nonumber \\
&\times& \exp \left\{ -\frac 1{2D_{xy}}\,\left[ \langle y^2\rangle (x-\langle
x\rangle )^2+\langle x^2\rangle (y-\langle y\rangle )^2-2\langle xy\rangle
(x-\langle x\rangle )(y-\langle y\rangle )\right] \right\} ,
\label{eq:CHI}
\end{eqnarray}
where 
\begin{eqnarray}
D_{xy}=\langle x^2\rangle \langle y^2\rangle -\langle xy\rangle
^2\,\,\,\,\,\,\,\,\,\,\,\,\,\,\,\,\,\,\,\,;\,\,\,\,\,\,\,\,\,\,\,\,\,\,\,\,%
\,\,\,\,\langle x^ny^m\rangle
=\int_0^{x_{max}}\!dx'\,x'^n\,\int_0^{y_{max}}\!dy'\,y'^m\,\omega (x',y'),
\label{eq:defMOM}
\end{eqnarray}
with $\chi _0$ being a normalization factor defined by the condition that 
\begin{eqnarray}
\int_0^1\!dx\,\int_0^1\!dy\,\chi (x,y)\theta (xy-K_{min}^2)=1
\end{eqnarray}
with $K_{min}=\frac{m_0}{\sqrt{s}}$ being the minimal inelasticity defined
by the mass $m_0$ of the lightest possible central gluonic cluster.

As it can be seen from (\ref{eq:CHI}),  the probability that the incoming hadrons release an energy of 
$\sqrt{x y s}$ is a two-dimensional gaussian function of $x$ and $y$ with maximum governed by 
the momenta $\langle x \rangle$ and $\langle y \rangle$ (particular cases of (\ref{eq:defMOM})), which, in turn, 
depend on the integration limits $x_{max}$ and  $y_{max}$.  
By reducing these maximal value we select events in which the energy released by the proton and by the 
antiproton is small (i.e., $M_X$ is small) and at the same time two rapidity gaps will be formed. This 
is how we define our ``kinematical Pomeron'': a set of gluons belonging to the proton (or antiproton) carrying 
altogether a small fraction of the parent hadron momentum. The function $\omega (x',y')$ will be discussed below.
In the formulation of the IGM $x_{max} = y_{max}=1$ if only 
non-diffractive processes are present. In the model, diffraction (double Pomeron exchange) means reducing $y_{max}$
($x_{max}$ and $y_{max}$). Although this procedure is somewhat arbitrary and we could choose any small number for
the integration limits, this freedom of choice is dramatically reduced if we use single diffractive events (SPE) as 
a guide. In \cite{IGM97} we have shown that the choice leading to the best description of diffractive mass 
spectra is $y_{max}=y$. Actually, using this cut in (\ref{eq:defMOM}) and some simple approximations (described 
in \cite{IGM97} and also in the appendix) we could  obtain the analytical formula for the single diffractive mass 
spectrum: 
\begin{eqnarray}
\frac{dN_{SPE}}{dM_X^2}\, &=&\,
\frac{\chi_0}{\pi\sqrt{c}}\frac{1}{
M_X^2}\frac{1}{[\ln\left(\frac{M_X^2}{m_0^2}\right)]^{1/2}}  
\exp\left\{ -\, \frac{\left[1\, -\,
c\ln\left(\frac{M_X^2}{m_0^2}\right)\right]^2}{c\ln
\left(\frac{M_X^2}{m_0^2}\right)}\right\} 
\label{analsd}
\end{eqnarray}
where $c$ is a constant (discussed below). This spectrum shape is in very good agreement with a wide body of data. 
Based on our previous success we shall  assume here that in double Pomeron exchange we have  $x_{max} = x$ and 
$y_{max}=y$ and consequently $x_{max} \,  y_{max} \, = x y \, =  \frac{M_x^2}{s}$. 

In \cite{gou} it has been conjectured that the ratio of two-gap to one-gap rates could be used to test QCD aspects 
of gap formation. We therefore calculate in the appendix, with the same approximations, the analytical expression 
for the DPE  mass spectrum, which turns out to be: 
\begin{eqnarray}
\frac{dN_{DPE}}{dM_X^2}\,  &=&\, \frac{\chi_0^{'}}{\pi c'} \frac{1}{M_X^2} \, 
\frac{\ln\left(\frac{s}{M_X^2}\right)}{\ln\left(\frac{M_X^2}{m_0^2}\right)} \,
\exp\left\{ -\, \frac{2\cdot \left[1\, -\,
c'\ln\left(\frac{M_X^2}{m_0^2}\right)\right]^2}{c'\ln
\left(\frac{M_X^2}{m_0^2}\right)}\right\} 
\label{analdd}
\end{eqnarray}

Both expressions (\ref{analsd}) and (\ref{analdd}) are dominated by the $1/M^2_X$ factor. This garantees a 
pronounced fall with increasing diffractive masses, which is confirmed by the numerical calculations 
(which have no approximation) presented below.

\subsection{Dynamics}

The spectral function, $\omega (x',y')$, 
contains all the dynamical input of the IGM. Their soft and semihard components are given by (cf. \cite{DNW}): 
\begin{equation}
\omega^S (x',y')\,=\,\frac{\hat{\sigma}^S _{gg}(x'y's)}{\sigma (s)}\,G(x')\,G(y')\,\theta
\left( x'y'-K_{min}^2\right) ,  \label{eq:OMEGAS}
\end{equation}
\begin{equation}
\omega^H (x',y')\,=\,\frac{\hat{\sigma}^H _{gg}(x'y's)}{\sigma (s)}\,G(x')\,G(y')\,\theta
\left( x'y'-\frac {4p_{T_{min}}^2}{s}\right) ,  \label{eq:OMEGAH}
\end{equation}
where $G$'s denote the effective number of gluons from the corresponding projectiles (approximated by the 
respective gluonic structure functions), $\hat{\sigma}^S _{gg}$ and  $\hat{\sigma}^H _{gg}$ are the soft 
and semihard gluonic cross sections, $p_{T_{min}}$ is the minimum transverse momentum for minijet production 
and $\sigma = \sigma_{I\!\!P I\!\!P}$ denotes the Pomeron-Pomeron cross section.

In order to be more precise, the function $G(x')$ ($G(y')$) represents the momentum distribution of the gluons 
belonging to the proton (antiproton) subset called Pomeron 
and $x'$ ($y'$) is the momentum fraction {\it of the proton} ({\it antiproton}) carried by one of these gluons. 
We shall therefore use the notation $ G(x') = G_{I\!\!P}(x')$. This  function should not be confused with the 
momentum distribution of the gluons inside the Pomeron, $f_{g/I\!\!P}(\beta)$. 

The Pomeron for us is just a collection of gluons which belong to the diffracted proton 
(antiproton). In our previous works we have assumed that these gluons behave like all other ordinary gluons 
in the proton and have therefore the same momentum distribution. The only difference is the momentum sum rule, 
which for the gluons in $I\!\!P$  is:
\begin{eqnarray}
\int_0^1\! dx'\,x'\,G_{I\!\!P}(x')\,=\,p_d
\label{sumrule}
\end{eqnarray}
where $p_d=0.05$ (see \cite{IGM97,LEADING})
instead of $p=0.5$, which holds for the entire gluon population in the proton.

In order to make 
contact with the analysis performed by HERA experimental groups we consider 
two possible momentum distributions for the gluons inside  $I\!\!P$. A hard 
one:
\begin{eqnarray}
f^h_{g/I\!\!P}(\beta) &=&  a_{h}\,(1\,-\,\beta) 
\label{fhard}
\end{eqnarray}
and a ``super-hard'' (as it is called in \cite{terron}) or ``leading gluon'' 
(as it is called in \cite{newzeus})  one:
\begin{eqnarray}
f^{sh}_{g/I\!\!P}(\beta) &=&  a_{sh}\, \beta^7 \, (1\,-\,\beta)^{0.3} 
\label{superhard}
\end{eqnarray}
where $\beta$ is the momentum fraction of the Pomeron carried by the gluons
and the superscripts $h$ and $sh$ denote hard and superhard respectively. The constants  
$a_{h}$ and $a_{sh}$ will be fixed by the sum rule (\ref{sumrule}). 
In the past \cite{hera}, following the same formalism, we have also considered 
a soft gluon distribution for the Pomeron of the type
\begin{equation}
f^s_{g/I\!\!P}(\beta) =  6\,\frac{(1\,-\,\beta)^5}{\beta}
\label{supersoft}
\end{equation}
but  we found that this ``soft Pomeron'' distribution was incompatible with 
the single diffractive mass spectra measured at HERA \cite{h197}. This Pomeron profile 
was also ruled out by other types of observables, as concluded in Refs. \cite{zeus95} and
\cite{h194}.

We shall use the Donnachie-Landshoff  Pomeron flux factor, which, after the integration in
the $t$ variable, is approximately given by \cite{terron}:
\begin{equation}
f_{I\!\!P/p}(x_{I\!\!P}) \simeq  C \, x_{I\!\!P}^{1-2 \alpha_ {I\!\!P}}  \simeq 
 C \,\frac{1}{x_{I\!\!P}} 
\end{equation} 
where $x_{I\!\!P}$ is the fraction of the proton momentum carried by the
Pomeron and the normalization constant $C$ will be fixed later, also with the help of 
(\ref{sumrule}). Noticing  that
$\beta = \frac{x}{x_{I\!\!P}}$ the distribution $G_{I\!\!P}(y)$ needed 
in eqs. (\ref{eq:OMEGAS}) and (\ref{eq:OMEGAH}) is then given by the convolution: 
\begin{eqnarray}
G^{h,sg}_{I\!\!P} (y) &=& \int_y^1\! \frac{ dx_{I\!\!P} } {x_{I\!\!P}} \,
f_{I\!\!P/p}(x_{I\!\!P})\,f^{h,sg}_{g/I\!\!P}(\frac{y}{x_{I\!\!P}})
\label{conv}
\end{eqnarray}

We shall use also the ``diffractive gluon distribution'' given by:
\begin{equation}
G_{I\!\!P}(y) =  a\,\frac{(1\,-\,y)^5}{y}
\label{gFeynman}
\end{equation}
where $a$ is fixed by the sum rule. With (\ref{gFeynman})
we could obtain a very good description of diffractive mass spectra \cite{IGM97,hera}. 
Therefore we shall keep using it here.

We are implicitly assuming that all gluons from $p$ and 
$\bar p$ participating in the collision (i.e., those emitted 
from the upper and lower vertex in Fig. 1) have to form a color singlet. Only then 
two large rapidity gaps will form separating the diffracted proton, 
the $M_X$ system and the diffracted antiproton, which 
is the experimental requirement defining a DPE event.

We can now calculate the diffractive mass distribution $M_X$ using the $\chi
(x,y)$ function by simply performing a change of variables: 
\begin{eqnarray}
\frac{1}{\sigma} \frac{d\sigma}{dM_X}\,&=&\,\int_0^1\!dx\,\int_0^1\!dy\,\chi (x,y)\,
\delta \left[M_X-\sqrt{xys}\right] \,\theta \left(xys-m_{0}^2 \right)\nonumber \\
&=&\frac {2\,M_X}{s}\,\int_{\frac {M_X^2}{s}}^1 \!dx\, \frac{1}{x}\,
\chi \left[ x,y = \frac{M_X^2}{xs}\right] \, \theta (M_X^2-m_0^2)
\label{eq:dndmx}
\end{eqnarray}

\section{Numerical Results and Discussion}

We start evaluating Eq. (\ref{eq:dndmx}) with the inputs that were already fixed by other applications of the 
IGM \cite{IGM97,hera}, namely, (\ref{gFeynman}) with $p_d=0.05$.  
In Fig. 2 we show the numerical  results for DPE mass distribution. 
We have fixed the parameter $\sigma$ ($\equiv \sigma_{I\!\!P I\!\!P}$) appearing in 
Eq. (\ref{eq:OMEGAS}) and Eq. (\ref{eq:OMEGAH}), to $0.5\,mb$ (solid lines) and $1.0\,mb$ (dashed lines). 
In both cases our curves were normalized to the  ``AND" (Fig. 2a) and ``OR"  (Fig. 2b) data samples of \cite{UA8}. 
We emphasize that, 
in this approach, since
we have fixed all parameters using previous data on leading particle formation and single diffractive 
mass spectra, there are no free parameters here, except $\sigma_{I\!\!P I\!\!P}$.

As one can see from the figures, in our model we obtain the fast increase of spectra in the low mass region 
without the use of a $M_X$ dependent  $I\!\!P$-$I\!\!P$ cross section and  this quantity seems to be 
approximately $\sigma_{I\!\!P I\!\!P} \simeq 0.5\,mb$. This is the main message of this 
note.

In order to  investigate the sensitivity of Eq. (\ref{eq:dndmx}) to the value of $p_d$, we show in 
Fig. 3 with arbitrary units, diffractive mass spectra obtained with $p_d= 0.025$ (dash-dotted line),  $0.05$ (solid line) 
and $0.1$ (dashed line). The three curves have the same normalization and we observe that, as $p_d$ increases 
our mass spectrum becomes softer. It is interesting to remark that, in the actual calculations, the parameter $p_d$ 
appears always divided by  $\sigma_{I\!\!P I\!\!P}$ in the computation of the moments (\ref{eq:defMOM}). In fact, 
they form one single parameter. Assuming that the Pomeron profile is universal, we could disentangle one from 
the other,  fitting $p_d$ from the analysis 
of previous data \cite{IGM97,hera} and now extracting $\sigma_{I\!\!P I\!\!P}$ from the the UA8 data. 

We next replace (\ref{gFeynman}) by the convolution (\ref{conv})  to see  which of the previously considered 
Pomeron profiles, hard or superhard, gives the best fit of the UA8 data. In doing so,  we shall keep everything else
the same, i.e., $p_d= 0.05$ and  $\sigma_{I\!\!P I\!\!P} = 0.5 \, mb$. In Fig.  4a we compute (\ref{conv})  and 
show  $ y \,G_{I\!\!P}(y) $ for  Eq. (\ref{gFeynman}) (solid line), the hard distribution (\ref{fhard}) (dashed line) 
and the superhard (\ref{superhard}) (dash-dotted line). For the sake of comparison,  Fig. 4b shows the corresponding 
diffractive mass spectra normalized to the unity with same 
notation. We see that, for harder Pomeron profiles we ``dig a hole'' in the low mass region of the spectrum.

In Fig. 5, we repeat the fitting procedure used in Fig. 2 for the Pomeron profiles shown in Fig. 4. We fix
$p_d=0.05$, $\sigma_{I\!\!P I\!\!P} = 0.5 \, mb$. Solid, dashed and dash-dotted lines represent respectively 
Eq. (\ref{gFeynman}), hard and superhard Pomerons. Note that the  solid lines are the same as in Fig. 2.  
Looking at the figure, at first sight, we might be tempted to say that Eq. 
(\ref{gFeynman}) gives the best agreement with data and a somewhat worse description can be obtained with the 
hard Pomeron (in dashed lines), the superhard being discarded.  However, comparing the dashed lines in Fig. 2  
and Fig. 5  and observing that they practically coincide with each other,  we conclude that the same curve can be 
obtained either with   (\ref{gFeynman}) and $\sigma_{I\!\!P I\!\!P} = 1.0 \, mb$ (dashed line in Fig. 2) or with 
(\ref{conv}), (\ref{fhard}) and $\sigma_{I\!\!P I\!\!P} = 0.5 \, mb$ (dashed line in Fig. 5). {\it In other words we 
can trade the ``hardness'' of the Pomeron with its interaction cross section. The following two objects give an 
equally good description of data: i) a Pomeron composed by more and softer gluons and with a larger cross 
section and ii) a Pomeron made by fewer, harder gluons with a smaller interaction cross section.} We have checked that
this reasoning can be extended to the superhard Pomeron. Although, apparently disfavoured by Fig. 5 (dash-dotted lines), 
it might still fit the data provided that $\sigma_{I\!\!P I\!\!P} < 0.25 \, mb$.  Given the uncertainties in the data 
and the limitations of the model,  we will not try for the moment to refine this analysis. It seems possible to 
describe data  in  a number of different ways. We conclude then that nothing exotic has been observed and also that 
the Pomeron-Pomeron cross section is bounded to be  $\sigma_{I\!\!P I\!\!P} < 1.0 \, mb$.

In Fig. 6a we compare our predictions for $d\sigma/dM_X \, (mb/GeV)$ at Tevatron  
($\sqrt s = 2 \, TeV$) and in Fig. 6b for LHC ($\sqrt s = 14 \, TeV$)  assuming an $M_X$-independent 
$\sigma_{I\!\!P I\!\!P}=1.0\,mb$  (and using (\ref{gFeynman})) with predictions made by Brandt {\it et al.} \cite{UA8} 
for two values of effective  Pomeron  intercepts ($\alpha (0) = 1+\varepsilon$),  $\varepsilon = 0.0 $ 
and $0.035$.

Although the normalization of our curves is arbitrary, the comparison of the shapes reveals a striking difference 
between the two predictions. Whereas the points (from \cite{UA8}) show spectra broadening with the c.m.s. energy, 
we predict (solid lines) the  opposite behavior:  as the energy increases we observe a (modest) narrowing for 
$d\sigma/dM_X$. This
small effect means that the diffractive mass becomes a smaller fraction
of the available energy $\sqrt{s}$. In other words, the ``diffractive
inelasticity" decreases with energy and consequently the ``diffracted
leading particles" follow a harder $x_F$ spectrum. Physically, in
the context of the IGM, this means that the deposited energy is
increasing with $\sqrt{s}$ (due the minijets) but it will be mostly released outside the 
phase space region that we are selecting. 

We are not able to make precise statements about the diffractive cross section (in particular about
its normalization)  with our simple model. Nevertheless, the narrowing of $d\sigma_{DPE}/dM_X$ suggests a 
slower increase  (with $\sqrt{s}$) of the integrated distribution $\sigma_{DPE}$. We found this same effect 
\cite{IGM97} also for $\sigma_{SPE}$. This trend is welcome and is one of the possible mechanisms responsible 
for the suppression  of diffractive cross sections at higher energies relative to Regge theory predictions.

In Fig. 7 we show the ratio $R(M_X)$ defined by:
\begin{equation}
R(M_X)= \frac{ \frac{1}{\sigma_{DPE}} \, \frac{d \sigma_{DPE}}{d M_X} } 
          {\frac{1}{\sigma_{SPE}} \, \frac{d \sigma_{SPE}}{d M_X}}
\label{ratioR}
\end{equation}
This quantity involves only distributions previously normalized to unity and does not directly compare the
cross sections (which are numerically very different for DPE and single diffraction). In $R$ the dominant 
$1/M^2_X$ factors cancel, as suggested by the comparison between (\ref{analsd}) and (\ref{analdd}),  and we 
can  better analyse the details of the distributions which may contain interesting dynamical information. 
The most prominent feature of Fig. 7 is the rise of the ratio with $M_X$, almost by one order of magnitude in 
the mass range considered.  This can be qualitatively attributed to the fact that, in single diffractive events 
the object  $X$ has larger rapidities than the corresponding cluster formed in DPE events. 
As a consequence, when energy is released from the incoming particles in a SPE event, it 
goes more to kinetic energy of the $X$ system (i.e., larger momentum $P_X$ and rapidity $Y_X$) and less to its 
mass.  In DPE, although less energy is released, it goes predominantly to the mass $M_X$ of
the difractive cluster, which is then at lower values of $Y_X$. In order to illustrate this behavior, we show in Fig. 8
the rapidity distributions of the $X$ system (which has $M_X$). All curves are normalized to unity and with 
them we just want to draw attention to the dramatically  different positions of the maxima of these distributions. 
The solid and dashed lines show $ 1/\sigma \, d \sigma / d Y_X $ for DPE (curves on the left) and SPE (curves on the right) 
computed at $\sqrt{s} = 630\,GeV$ and $\sqrt{s} = 2000\,GeV$, respectively. We can clearly observe 
that DPE and SPE rapidity distributions are separated by three units of rapidity and this difference stays 
nearly constant as the c.m.s. energy increases. The location of maxima in $ 1/\sigma \, d \sigma / d Y_X $ and 
their energy dependence are predictions of our model.

To summarize: we have further developed our model for hadronic collisions and included double Pomeron exchange 
events.
With only one new parameter, $\sigma_{I\!\!P I\!\!P}$, we could fit the data recently published by the UA8 
collaboration and make predictions for the DPE mass spectra at Tevatron and LHC energies. Our 
main conclusion is that $\sigma_{I\!\!P I\!\!P} \simeq 0.5\,mb$ and constant with $M_X$ is favored by experimental 
data. We predict that the ratio between (normalized) double $I\!\!P$-$I\!\!P$ exchange and single diffractive mass 
distributions grows with $M_X$.  

\section{Appendix}

In what follows we shall often make use of our kinematical constraint between $x_m (\equiv x_{max})$ and 
$y_m (\equiv y_{max})$:

\begin{equation}
x_m y_m\, = x y \, =  \frac{M_X^2}{s}
\end{equation}

The moments of $\omega^S (x',y')$ and $\omega^H (x',y')$ are given by:
\begin{equation}
\langle x^ny^m\rangle_S =\int_0^{x_{m}}\!dx'\,x'^n\,\int_0^{y_{m}}\!dy'\,y'^m\,\omega^S (x',y')
=\int^x_{\frac{x m^2_0}{M^2_X}}\!dx'\,x'^n\,\int_{\frac {m^2_0}{x's}}^{\frac {M^2_X}{xs}}\!dy'\,y'^m\,
\omega^S (x',y')
\label{momegas}
\end{equation}
\begin{equation}
\langle x^ny^m\rangle_H =\int_0^{x_{m}}\!dx'\,x'^n\,\int_0^{y_{m}}\!dy'\,y'^m\,\omega^H (x',y')
=\int^x_{\frac {4 x p_{T_{min}}^2}{M^2_X}}\!dx'\,x'^n\,\int_{\frac {4 p_{T_{min}}^2}{x's}}
^{\frac {M^2_X}{xs}}\!dy'\,y'^m\,\omega^H (x',y')
\label{momegah}
\end{equation}

In order to obtain analytical expressions we shall, in the following set $\omega^H (x',y') = 0$ because 
at the relevant energies hard processes are not yet dominant. We shall keep only the low $x$ dominating 
factor of the gluon distribution:
\begin{equation}
G_{I\!\!P}(x')\, =  \, \frac {1}{x'}
\end{equation}
We shall neglect the $x'$ and $y'$ dependence of the cross sections and assume that:
\begin{equation}
\frac{ p_d^2 \,\sigma_{gg}}{\sigma_{I\!\!P I\!\!P}} \, = \, c
\end{equation}
With all these approximations (\ref{momegas}) can be rewritten as:
\begin{eqnarray}
\langle x^n\, y^m\rangle\, &=&\, \int_0^{x_m}\, dx'\, x'^n\,
\int_0^{y_m}\, dy'\, y'^m\, \omega(x',y')\, \theta\left(x'y' -
\frac{m_0^2}{s}\right) \nonumber \\
&=&\, c\, \int_0^{x_m}\, dx'\, x'^{n-1}\, \int_0^{y_m}\, dy'\, y'^{m-1}\, 
\theta\left(x'y' - \frac{m_0^2}{s}\right) \nonumber \\
&=&\, c\, \int_{\frac{m_0^2}{sy_m}}^{x_m}\, dx'\, x'^{n-1}\,
\int_{\frac{m_0^2}{sx}}^{y_m}\, dy'\, y'^{m-1} .
\end{eqnarray}

\subsection{ Case n=1, m=0}

\begin{eqnarray}
\langle x\rangle\, &=&\, c\, \int_{\frac{m_0^2}{sy_m}}^{x_m}\, dx'\, 
\int_{\frac{m_0^2}{sx}}^{y_m}\, \frac{dy'}{y'}\, =\, c\,
\int_{\frac{m_0^2}{sy_m}}^{x_m}\, dx'\,
\ln\left(\frac{y_ms}{m_0^2}\cdot x'\right) \nonumber\\
&=&\, c\cdot \frac{m_0^2}{y_ms}\cdot
\left[\left(\frac{M_X^2}{m_0^2}\right)\,
\ln\left(\frac{M_X^2}{m_0^2}\right)\, -\,
\left(\frac{M_X^2}{m_0^2}\right)\, +\, 1\, \right]\nonumber\\
&\simeq&\,  c\cdot \left(\frac{M_X^2}{s}\right) \cdot
\frac{1}{y_m}\cdot \ln \left(\frac{M_X^2}{m_0^2}\right) \, =\, c\cdot
x_m\cdot \ln \left(\frac{M_X^2}{m_0^2}\right) 
\end{eqnarray}

By symmetry we have:
\begin{equation}
\langle y\rangle \, \simeq \, c\cdot \left(\frac{M_X^2}{s}\right) \cdot
\frac{1}{x_m}\cdot \ln \left(\frac{M_X^2}{m_0^2}\right) \, =\, c\cdot
y_m\cdot \ln \left(\frac{M_X^2}{m_0^2}\right) 
\end{equation}

\subsection{ Case n=2, m=0}

\begin{eqnarray}
\langle x^2\rangle\, &=&\, c\, \int_{\frac{m_0^2}{sy_m}}^{x_m}\, dx'\, x'\,
\int_{\frac{m_0^2}{sx}}^{y_m}\, \frac{dy'}{y'}\, =\, c\,
\int_{\frac{m_0^2}{sy_m}}^{x_m}\, dx'\, x'\,
\ln\left(\frac{y_ms}{m_0^2}\cdot x\right) \nonumber\\
&\simeq&\, c\cdot \left(\frac{m_0^2}{y_ms}\right)^2 \cdot \frac{1}{2}
\left(\frac{M_X^2}{m_0^2}\right)^2 \ln \left(\frac{M_X^2}{m_0^2}\right) \nonumber\\
&\simeq& \,
\frac{1}{2}\left(\frac{M_X^2}{s}\right)\cdot \frac{1}{y_m}\cdot \langle
y\rangle \, =\, \frac{1}{2} \cdot x_m \cdot \langle x\rangle 
\end{eqnarray}

Again, by symmetry, we have:

\begin{equation}
\langle y^2\rangle\, \simeq\, =\, 
\frac{1}{2}\left(\frac{M_X^2}{s}\right)\cdot \frac{1}{x_m}\cdot \langle
y\rangle \, =\, \frac{1}{2} \cdot y_m \cdot \langle y\rangle 
\end{equation}

\subsection{ Case n=1, m=1}

\begin{eqnarray}
\langle xy\rangle\, &=&\, c\, \int_{\frac{m_0^2}{sy_m}}^{x_m}\, dx'\, 
\int_{\frac{m_0^2}{sx}}^{y_m}\, dy'\, =\, c\,
\int_{\frac{m_0^2}{sy_m}}^{x_m}\, dx'\, \left[ y_m\, -\,
\frac{m_0^2}{sx'} \right] \nonumber\\
&=&\, c\, \left[\left(x_my_m\right)\, -\, \frac{m_0^2}{s}\right]\,
-\, c\cdot\frac{m_0^2}{s}\ln\left( \frac{x_my_m\cdot s}{m_0^2}\right)\,
\simeq c\cdot \left(\frac{M_X^2}{s}\right)\, \simeq\, 0
\end{eqnarray}

Inserting the approximate expressions for the moments into (\ref{eq:CHI}) we
obtain:
\begin{eqnarray}
\chi(x,y)\, &\simeq&\, \frac{\chi_0}{2\pi\sqrt{D_{xy}}}\cdot 
  \exp \left\{ - \frac{\langle y^2\rangle (x - \langle x\rangle )^2 +
  \langle x^2\rangle (y - \langle y\rangle )^2}{2D_{xy}}\right\} , \\
D_{xy}\, &\simeq&\, \langle x^2\rangle \langle y^2\rangle 
\end{eqnarray}

or

\begin{equation}
\chi(x,y)\, \simeq\, \frac{\chi_0}{2\pi\sqrt{\langle
                    x^2\rangle\langle y^2\rangle}}\cdot 
                 \exp \left[ -\, \frac{(x - \langle x\rangle
)^2}{2\langle x^2\rangle} \,  
               -\, \frac{(y - \langle y\rangle )^2}{2\langle
y^2\rangle}\right]. 
\end{equation}

The diffractive mass distribution will be given by:

\begin{eqnarray}
\frac{dN}{dM_X^2}\, &=&\, \int_0^1\! dx\, \int_0^1\! dy\, 
                       \chi(x,y)\, \delta\left( M^2_X - xys\right)\, 
                       \theta \left( xy - \frac{m_0^2}{s}\right)
                       \nonumber \\
                    &=& 
          \frac{1}{s}\, \int_{_{\frac{M_X^2}{s}}}^1\! \frac{dx}{x}\,
          \chi\left( x,\frac{M_X^2}{x s}\right), \label{eq:RES}
\end{eqnarray}                       

In order to proceed further let us first notice that $x_m$ and $y_m$ in the 
formulas for $<x^ny^m>$ above will have the  meaning
of the $ x$ and $y$ because of the $\delta(M_X^2-xys)$
constraint. Therefore:

\begin{eqnarray}
\langle x\rangle = a\cdot \frac{1}{y_m}\, =\, \frac{a}{y} \, \, \, \, \, \, \,;\, \, \, \, \, \, \,
\langle y\rangle = a\cdot \frac{1}{x_m}\, =\, \frac{a}{x} 
\end{eqnarray}

\begin{eqnarray}
\langle x^2\rangle =
\frac{1}{2}\frac{a}{y}\cdot\left(\frac{M_X^2}{s}\right)\cdot
\frac{1}{y}\, =\, \frac{b}{y^2} \, \, \, \, \, \, \,;\, \, \, \, \, \, \,
\langle y^2\rangle =
\frac{1}{2}\frac{a}{x}\cdot\left(\frac{M_X^2}{s}\right)\cdot
\frac{1}{x}\, =\, \frac{b}{x^2}  
\end{eqnarray}
where
\begin{equation}
a\, =\, c\cdot \left(\frac{M_X^2}{s}\right)
\ln\left(\frac{M_X^2}{m_0^2}\right) \, \, \, \, \, \, \,;\, \, \, \, \, \, \,
b\, =\, \frac{1}{2}a\left(\frac{M_X^2}{s}\right) 
\end{equation}

It then follows that:

\begin{eqnarray}
&&\exp \left[ -\, \frac{(x - \langle x\rangle )^2}{2\langle
x^2\rangle} \, -\, \frac{(y - \langle y\rangle )^2}{2\langle
y^2\rangle}\right]\, =\, \nonumber\\
&&\exp \left[ -\, \frac{\left(x - \frac{a}{y}\right)^2}
                       {2\frac{b}{y^2}}
              -\,     \frac{\left(y - \frac{a}{x}\right)^2}
                       {2\frac{b}{x^2}}\right]\, =\, 
  \exp\left[ -\,\frac{(xy\, -\, a)^2}{b}\right] .
\end{eqnarray}

Using once again that $xy = \frac{M_X^2}{s}$
we arrive at:

\begin{eqnarray}
  \exp\left[ 
             -\, \frac{(xy\, -\, a)^2}{b} 
                                           \right]\, &=&\,
  \exp\left\{ 
  -\, \frac{
            2 \, \left[
                    \frac{M_X^2}{s} -
          c\left(\frac{M_X^2}{s}\right) \ln \left(\frac{M_X^2}{m_0^2}\right)
             \right]^2}
{c\left(\frac{M_X^2}{s}\right)^2 \ln \left(\frac{M_X^2}{m_0^2}\right)}
\right\} \nonumber\\
&=&\, \exp\left\{ -\, \frac{ 2 \,\left[1\, -\,
c\ln\left(\frac{M_X^2}{m_0^2}\right)\right]^2}{c\ln
\left(\frac{M_X^2}{m_0^2}\right)}\right\} 
\end{eqnarray}

The total $\chi(x,y)$ is then:

\begin{equation}
\chi(x,y)\, =\, \frac{\chi_0}{\pi c
\left(\frac{M_X^2}{s}\right)\ln\left(\frac{M_X^2}{m_0^2}\right)} 
\exp\left\{ -\, \frac{ 2 \, \left[1\, -\,
c\ln\left(\frac{M_X^2}{m_0^2}\right)\right]^2}{c\ln
\left(\frac{M_X^2}{m_0^2}\right)}\right\} 
\end{equation}

leading to

\begin{eqnarray}
\frac{dN}{dM_X^2}\,  &=&\, \frac{\chi_0}{\pi c} \frac{1}{M_X^2} \, 
\frac{\ln\left(\frac{s}{M_X^2}\right)}{\ln\left(\frac{M_X^2}{m_0^2}\right)} \,
\exp\left\{ -\, \frac{2\cdot \left[1\, -\,
c\ln\left(\frac{M_X^2}{m_0^2}\right)\right]^2}{c\ln
\left(\frac{M_X^2}{m_0^2}\right)}\right\} 
\end{eqnarray}

which is exactly Eq. (\ref{analdd}).

\underline{Acknowledgements}: This work has been supported by FAPESP, CNPQ (Brazil) and KBN (Poland).

\begin{figure} 
\begin{center}
\epsfysize=13.0cm
\centerline{\epsfig{figure=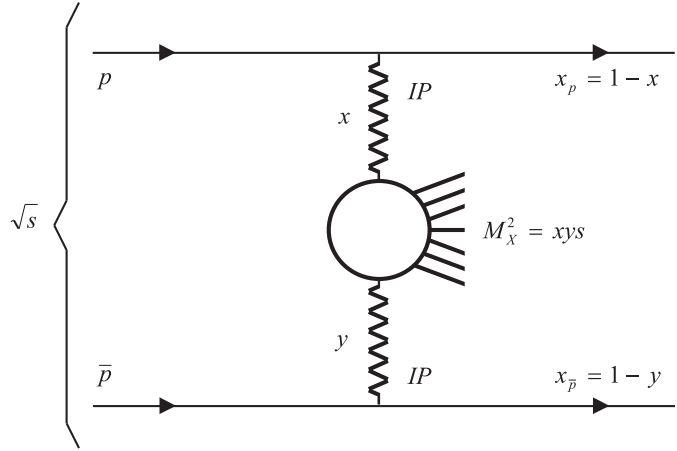,width=9.0cm}} 
\caption{IGM picture for a double Pomeron exchange process.}
\end{center}
\end{figure}

\begin{figure} 
\begin{center}
\centerline{\epsfig{figure=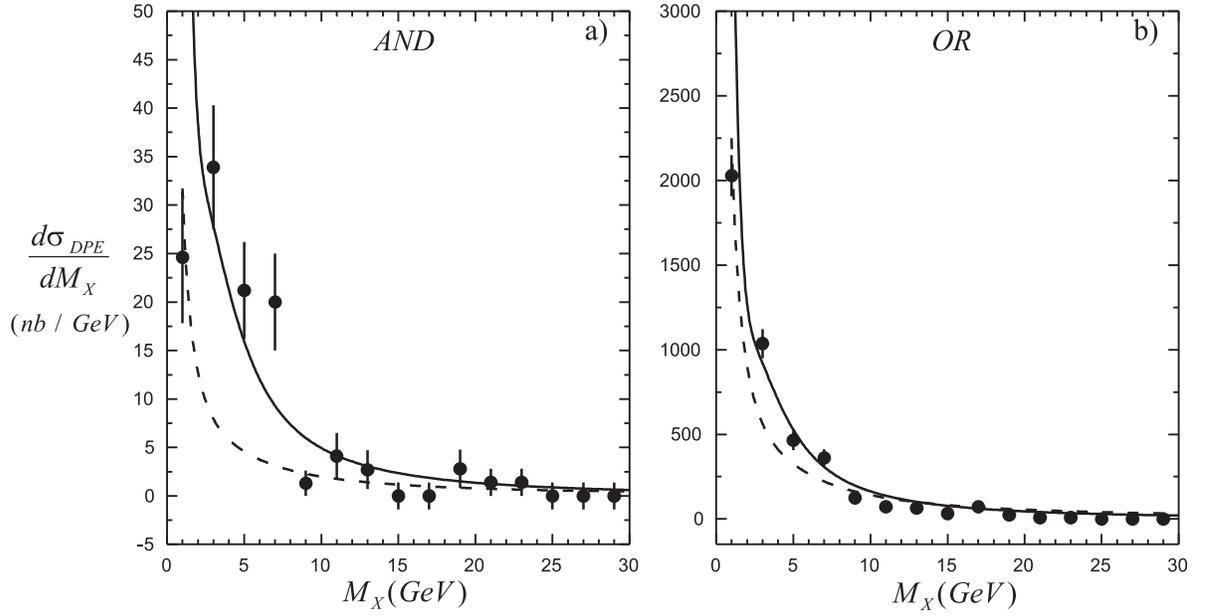,width=16.0cm}}
\caption{IGM DPE diffractive mass distributions: Solid and dashed lines show the numerical results with $\sigma_{I\!\!P I\!\!P}$  equal to $0.5\,mb$ and $1.0\,mb$, respectively. Our curves were normalized to the  ``AND" (a) and ``OR" (b) data samples of \protect\cite{UA8}.}
\end{center}
\end{figure}

\begin{figure} 
\begin{center}
\centerline{\epsfig{figure=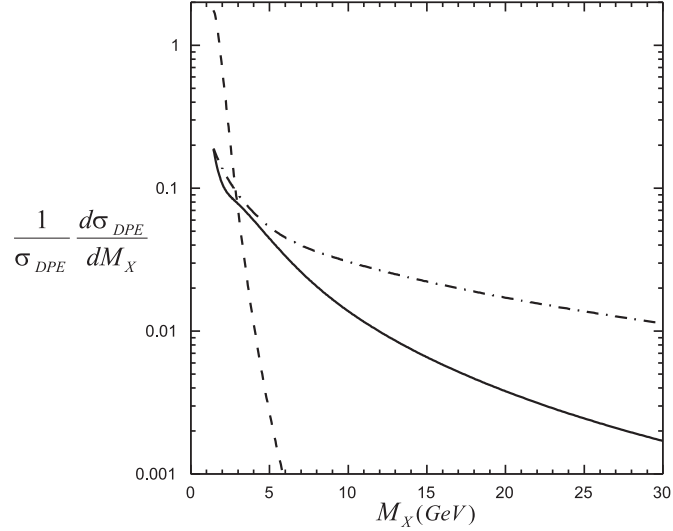,width=9.0cm}} 
\caption{IGM DPE diffractive mass distributions, in arbitrary units, with  $p_d= 0.025$ (dash-dotted line), $0.05$ (solid line) and $0.1$ (dashed line). In all cases  $\sigma_{I\!\!P I\!\!P}=0.5\,mb$.}
\end{center}
\end{figure}

\begin{figure} 
\begin{center}
\vskip -1cm
\centerline{\epsfig{figure=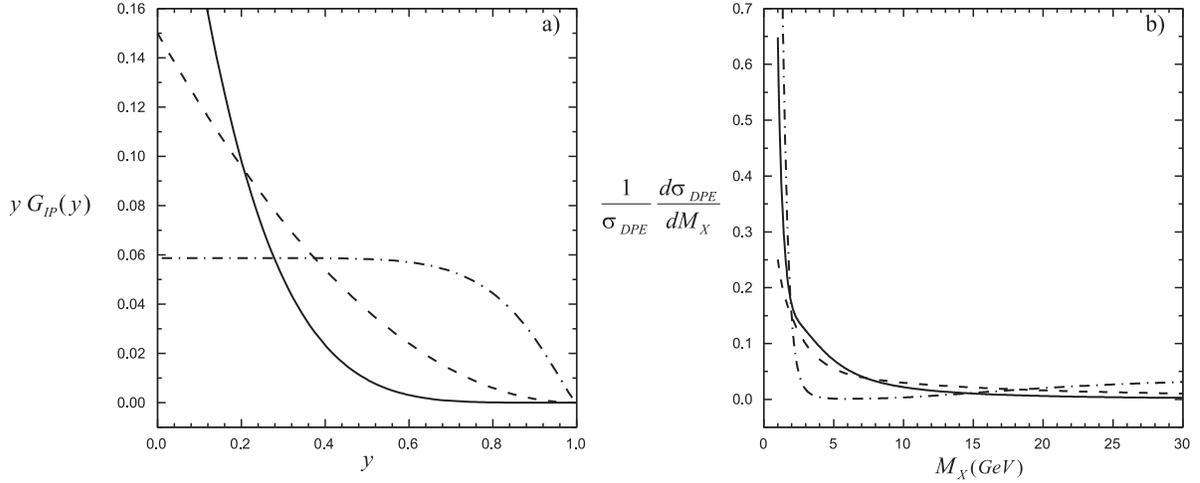,width=16.0cm}}
\caption{a) Diffractive gluon distributions: solid line was calculated with Eq. (\ref{gFeynman}), 
dashed and dash-dotted lines, calculated by Eq. (\ref{conv}), represent the ``hard" and ``super-hard" Pomeron profiles, given respectively by Eq. (\ref{fhard}) and (\ref{superhard}); 
b) Diffractive mass distributions, normalized to the unity, for the same cases showed in a). In all cases  $p_d= 0.05$ and $\sigma_{I\!\!P I\!\!P}=0.5\,mb$.}
\end{center}
\end{figure}

\begin{figure} 
\begin{center}
\centerline{\epsfig{figure=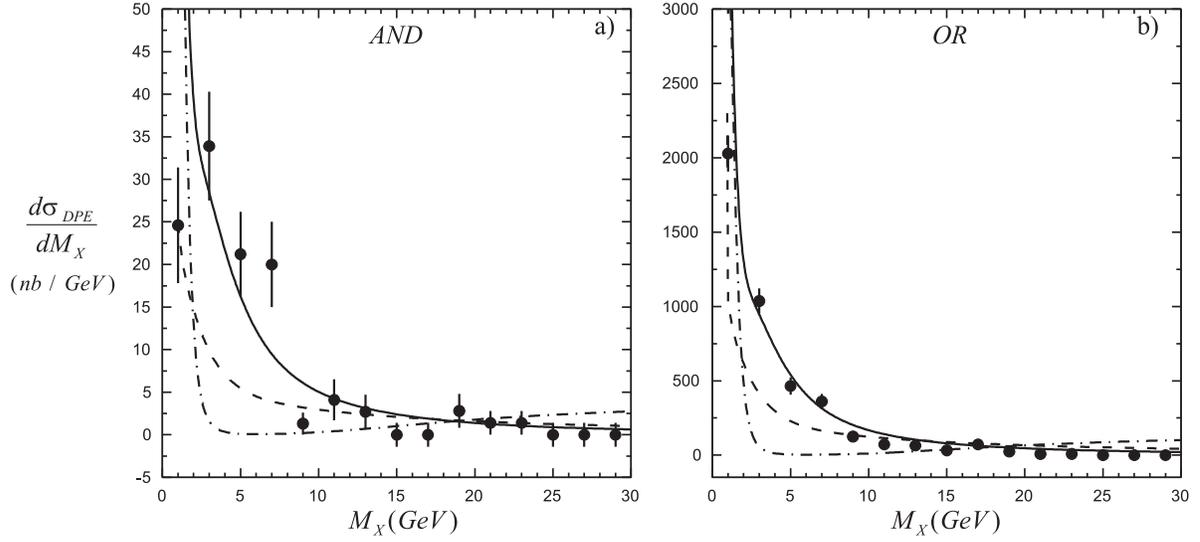,width=16.0cm}} 
\caption{IGM DPE diffractive mass distributions: Solid, dashed and dash-dotted lines follow the same notation of Fig. 4. 
Our curves were obtained with $p_d= 0.05$ and $\sigma_{I\!\!P I\!\!P}=0.5\,mb$ and normalized to the 
data samples of \protect\cite{UA8}.}
\end{center}
\end{figure}

\begin{figure} 
\begin{center}
\centerline{\epsfig{figure= 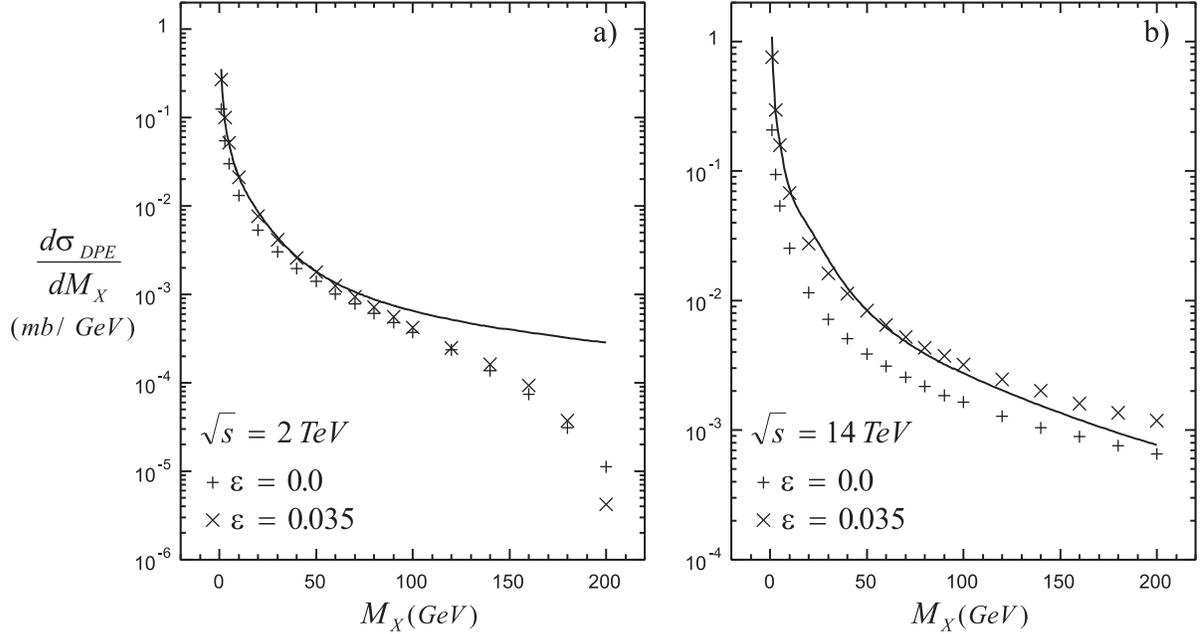,width=16.0cm}} 
\caption{IGM predictions for $d\sigma/dM_X$ at Tevatron and at LHC with $\sigma_{I\!\!P I\!\!P}=1.0\,mb$. Cross($+$) and Cross($\times$) are predictions made by Brandt {\it et al.} \protect\cite{UA8} for two values of effective Pomeron intercepts $\alpha (0) = 1+\varepsilon$.}
\end{center}
\end{figure}

\begin{figure} 
\begin{center}
\centerline{\epsfig{figure=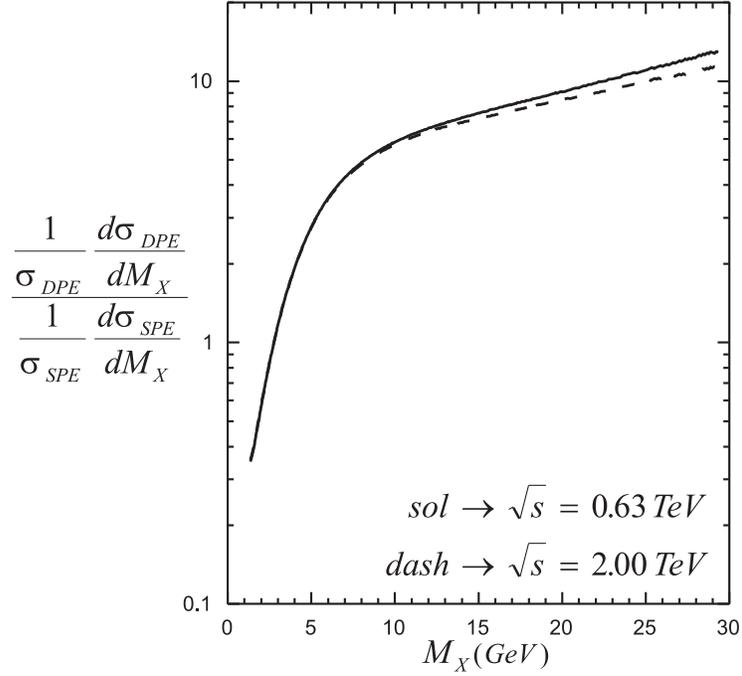,width=10.0cm}} 
\caption{Ratio double/single Pomeron exchange mass distributions as a function of $M_X$. In both cases we have assumed $\sigma_{I\!\!P I\!\!P}=1.0\,mb$ (for DPE processes) and $\sigma_{p I\!\!P}=1.0\,mb$ (for SPE processes).}
\end{center}
\end{figure}

\begin{figure} 
\begin{center}
\centerline{\epsfig{figure=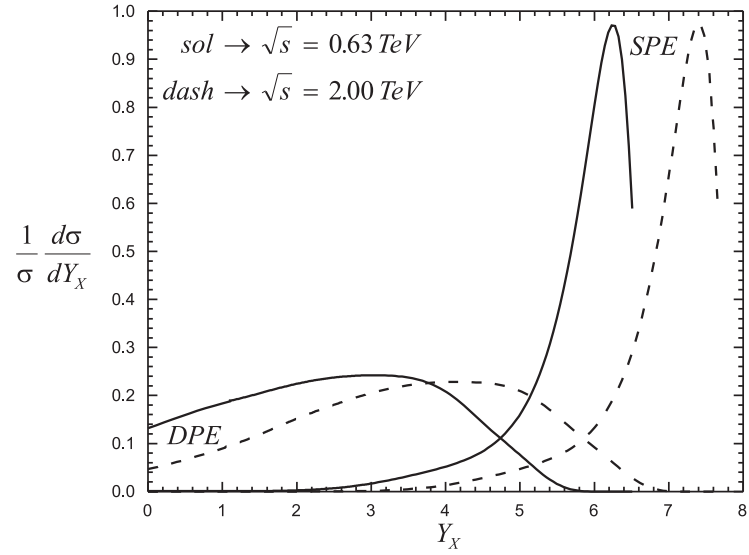,width=10.0cm}} 
\caption{Double and single Pomeron exchange normalized rapidity ($Y_X$) distributions. In both cases we have assumed $\sigma_{I\!\!P I\!\!P}=1.0\,mb$ (for DPE processes) and $\sigma_{p I\!\!P}=1.0\,mb$ (for SPE processes).}
\end{center}
\end{figure}

\end{document}